\documentclass[a4paper,12pt]{article}
\usepackage{graphicx} 
\usepackage{epstopdf}
\usepackage{ulem}
\usepackage{color}
\usepackage[numbers]{natbib}
\usepackage{float}
\usepackage{url}
\usepackage{amsmath} 
\usepackage{amssymb}
\usepackage{hyperref}
\renewcommand{\citep}[1]{[\citenum{#1}]}

\usepackage[labelfont=bf,labelsep=space]{caption}

\date{}
\begin{document}

\hoffset = -1truecm \voffset = -2truecm \baselineskip = 10 mm

\title{\bf{Gluon Condensation as a Unifying Mechanism for Special Spectra of Cosmic Gamma Rays and Low-Momentum Pion Enhancement at
the Large Hadron Collider}} 
\author{Wei Zhu$^{1,2,}$*, Jianhong Ruan$^{2}$, Xurong Chen$^{3}$, Yuchen Tang$^{4}$
\\\\
\normalsize $^{1}$ Center for Fundamental Physics and School of Mathematics and
Physics, \\
\normalsize Hubei Polytechnic University, Huangshi, 435003, China;\\
\normalsize $^{2}$ Department of Physics, East China Normal University, Shanghai,
200241, China;\\
\normalsize $^{3}$ Institute of Modern Physics, Chinese Academy of Sciences, Lanzhou,
730000, China; \\
\normalsize $^{4}$ College of science, Westlake University, Hangzhou, 310030, China;\\
\normalsize $^{*}$ Correspondence: wzhu@phy.ecnu.edu.cn}

\maketitle

\textbf{Abstract}
Decoding the internal structure of the proton is a fundamental challenge in physics. Historically, any new discovery about the proton has fuelled advances in several scientific fields. We have reported that gluons inside the proton accumulate near the critical momentum due to chaotic phenomena, forming gluon condensation. Surprisingly, the pion distribution predicted by this gluon distribution for the production of high-energy proton collisions could answer two puzzles in astronomy and high-energy physics. We find that during ultrahigh-energy cosmic ray collisions, gluon condensation may abruptly produce a large number of low-momentum pions, whose electromagnetic decays have the typical breakout properties appearing in various cosmic gamma-ray spectra.
   On the other hand, the Large Hadron Collider (LHC), which is well below the cosmic ray energy scale, also shows weak but recognisable signs of gluon condensation, which had been mistaken for BEC pions. The connection between these two phenomena, which occur at different scales in the Universe, supports the existence of a new structure within the proton-gluon condensation.

\textbf{Kerwords}
gluon condensation; Bose–Einstein condensation; pion condensate; color glass
condensate; relativistic heavy-ion collisions; cosmic gamma rays

\newpage

\section{Introduction}
The proton, a~fundamental building block of the universe, is composed of quarks and gluons. The~distribution of gluons has long posed a challenge in particle physics. Quantum chromodynamics (QCD) evolution equations predict that, at~high energies, gluon densities become extremely large. Under~these conditions, nonlinear effects dominate, giving rise to phenomena such as the color glass condensate (CGC) \citep{McLerran2011}. However, {CGC is not a true physical condensation. Further theoretical advancements introduced the Zhu--Shen--Ruan (ZSR) equation \citep{Zhu2016}. This equation indicates that, at~high energies, the~evolution of gluon distributions within nucleons exhibits chaotic behaviour. Such chaos leads to significant shadowing and antishadowing effects \citep{Zhu2016}. Consequently, a~large number of gluons accumulate within a narrow phase space defined by a critical momentum $(x_c, k_c)$. This phenomenon is known as gluon condensation (GC) \citep{Zhu2022}. Although~GC is consistent with the frameworks of the Standard Model and nonlinear science, its occurrence was unexpected by the scientific community. As~a result, identifying experimental evidence for GC has become our primary research objective.}

\begin{figure}[H]
    
\includegraphics[width=0.6\textwidth]{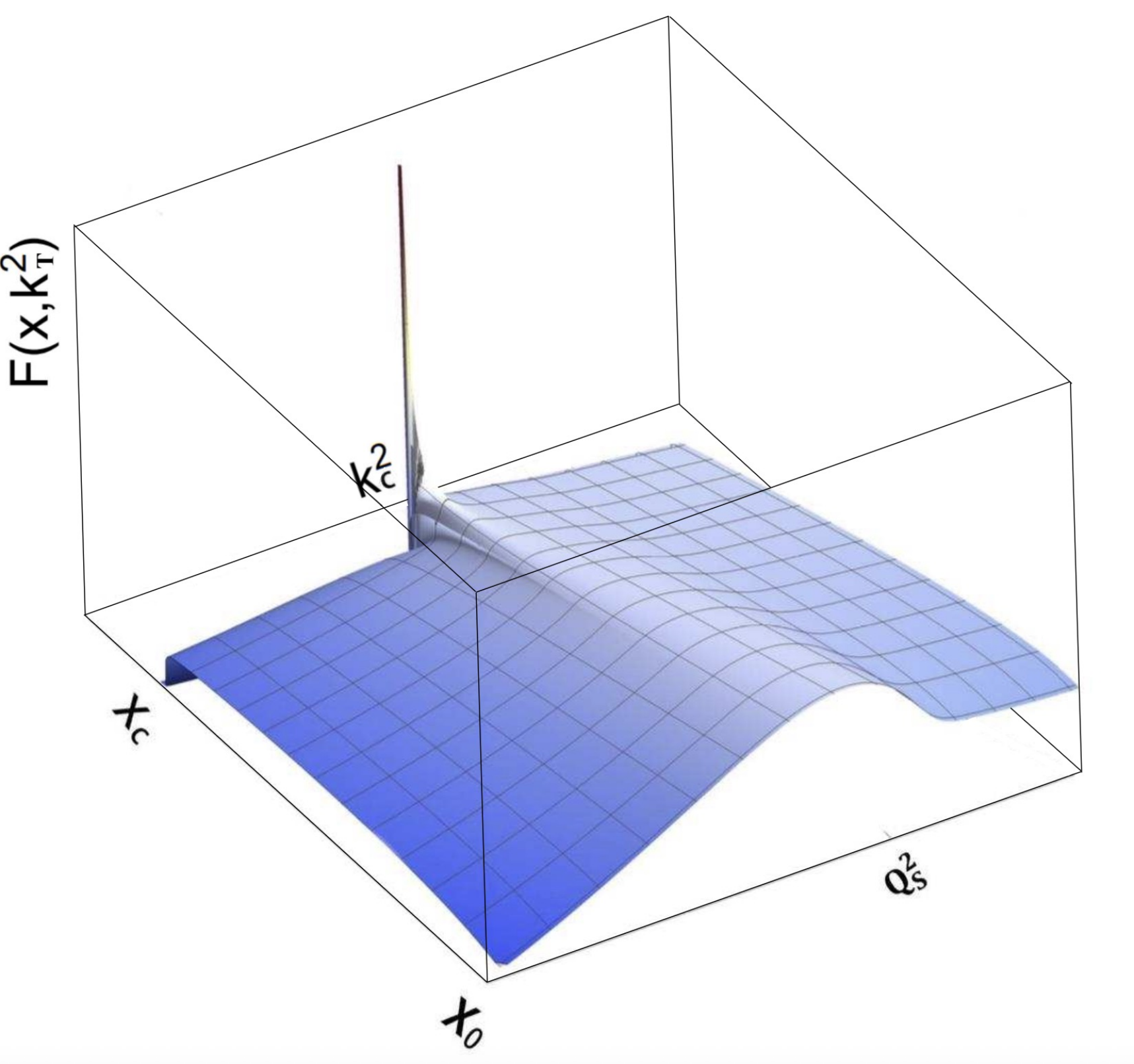}
\centering
    	   	\caption{The evolution of gluon distribution in a QCD evolution equation from a CGC model
    		to GC, where gluons at $x<x_c$ are condensed at a critical momentum $(x_c,k_c)$. All coordinates are on the
            logarithmic scale. There are two characteristics of this distribution: a sharp peak at the critical momentum and no gluons present at $x<x_c$.
    	}\label{fig:1}
\end{figure}

{The number of pions is expected to increase markedly when the dense peak of the gluon distribution in the proton enters the proton--proton ($p-p$) interaction region. Interestingly, by~incorporating the GC distribution (see~Figure
 \ref{fig:1}) into the general formula for proton collision cross sections, we can unify previously observed anomalies. These include features in the cosmic gamma-ray spectrum on an astronomical scale and in heavy-ion collisions at the Large Hadron Collider (LHC). This unification points to a previously unrecognized internal structure within the proton. We will detail this approach in Section~\ref{sec2}, where we present the relevant formulae used in this work. For~further information, please refer to the cited literature.}

 \vspace{-6pt}

In many astrophysical processes, protons can be accelerated to extremely high energies. When they collide with other protons or nuclei, they emit gamma rays, which may reveal information about the proton's interior. The broken power law (BPL) is a simple broken line in double logarithmic coordinates, often observed in high-energy gamma-ray spectra. There are many speculations and uncertainties about its formation \citep{Aharonian2012}. We will demonstrate that the GC model can provide a simple and unique explanation in Section~\ref{sec3}.

High-energy heavy-ion collisions have long sought the Bose--Einstein condensation (BEC) of pions. A~key feature is the large number of pions occupying low-momentum ($p\approx 0$) ground states. For~instance, Begun and Florkowski used the BEC model to fit LHC (ALICE) data, {finding pion coalescence to be 0--2\% in central collisions and rising to 19\% in peripheral collisions} \citep{Begun2015}. However, the~existence of BEC pions remains widely debated. In~Section~\ref{sec4}, we show that the GC model can produce similar results to Begun and Florkowski without the strict BEC~condition.

Thus, on~the deepest level of the proton---the very small $x$ gluon distribution---we have uncovered a possible intrinsic connection between the peculiar shape of the cosmic gamma-ray spectrum (at the astronomical scale) and the low momentum pion anomaly found at the LHC (at the particle scale). We will discuss them in the last~section.

\section{From CGC to~GC}\label{sec2}

     According to the GC model, GC is a result of QCD evolution from CGC \citep{Zhu2022}. The hadron collisions provide a means to observe this evolutionary process.
The production of secondary hadrons, primarily pions, through high-energy proton or nucleus collisions is divided into two steps at the quark--gluon level. This involves (i) the initial gluons from two hadron combine gluon mini-jets and (ii) the hadronization of mini-jets. Now we submit the GC distribution in Figure~\ref{fig:1} into a general formula for calculating the differential cross-section of gluon mini-jets \citep{Gribov1983,Szczurek2003}
\begin{equation}
    \frac{dN_g}{dk_T^2dy}=\frac{64N_c}{(N^2_c-1)k_T^2}\int q_T d
q_T\int_0^{2\pi}
d\phi\alpha_s(\Omega)\frac{F(x_1,\frac{1}{4}(k_T+q_T)^2)F(x_2,\frac{1}{4}
(k_T-q_T)^2)}{(k_T+q_T)^2(k_T-q_T)^2},
\label{eq:1}
\end{equation}
where
$\Omega=Max\{k_T^2,(k_T+q_T)^2/4, (k_T-q_T)^2/4\}$. The longitudinal
momentum fractions of interacting gluons are fixed by kinematics
$x_{1,2}=k_Te^{\pm y}/\sqrt{s}$; one can directly
obtain {the} following rapidity and transverse momentum distributions of gluon~mini-jets.

    Since $x\leq 1$, we have
\begin{equation}
    \vert y_{max}\vert=\ln \frac{\sqrt s}{k_{T,min}},
    \label{eq:2}
\end{equation}
and
\begin{equation}
    x_{1,2,min}=\frac{k_{T,min}}{\sqrt{s}}e^{-\vert y_{max}\vert}=\frac{k_{T,min}^2}{s}.
    \label{eq:3}
\end{equation}

We indicate that
the GC effects begin work from $\sqrt{s_{GC}}$ and finish at $\sqrt{s_{max}}$. Therefore,
\begin{equation}
    x_c\equiv \frac {k_c^2}{s_{GC}},
    \label{eq:4}
\end{equation}
and
\begin{equation}
    {x_c\equiv \frac{k_c}{\sqrt {s_{max}}}}.
    \label{eq:5}
\end{equation}

    We use the blue and black curves to describe the results from the GBW distribution \mbox{(a CGC model)} with and without
the GC effect, respectively. We were surprised to find the significant difference between two results, and~we propose that this difference can unify the anomalies in the cosmic gamma-ray spectrum (at astronomical scale) and heavy-ion collisions
(at particle scale) at the Large Hadron Collider (LHC) due to the~GC effects.

\section{The BPL in Cosmic Gamma Ray~Spectra}\label{sec3}

\begin{figure}[H]
    
\includegraphics[width=0.9\textwidth]{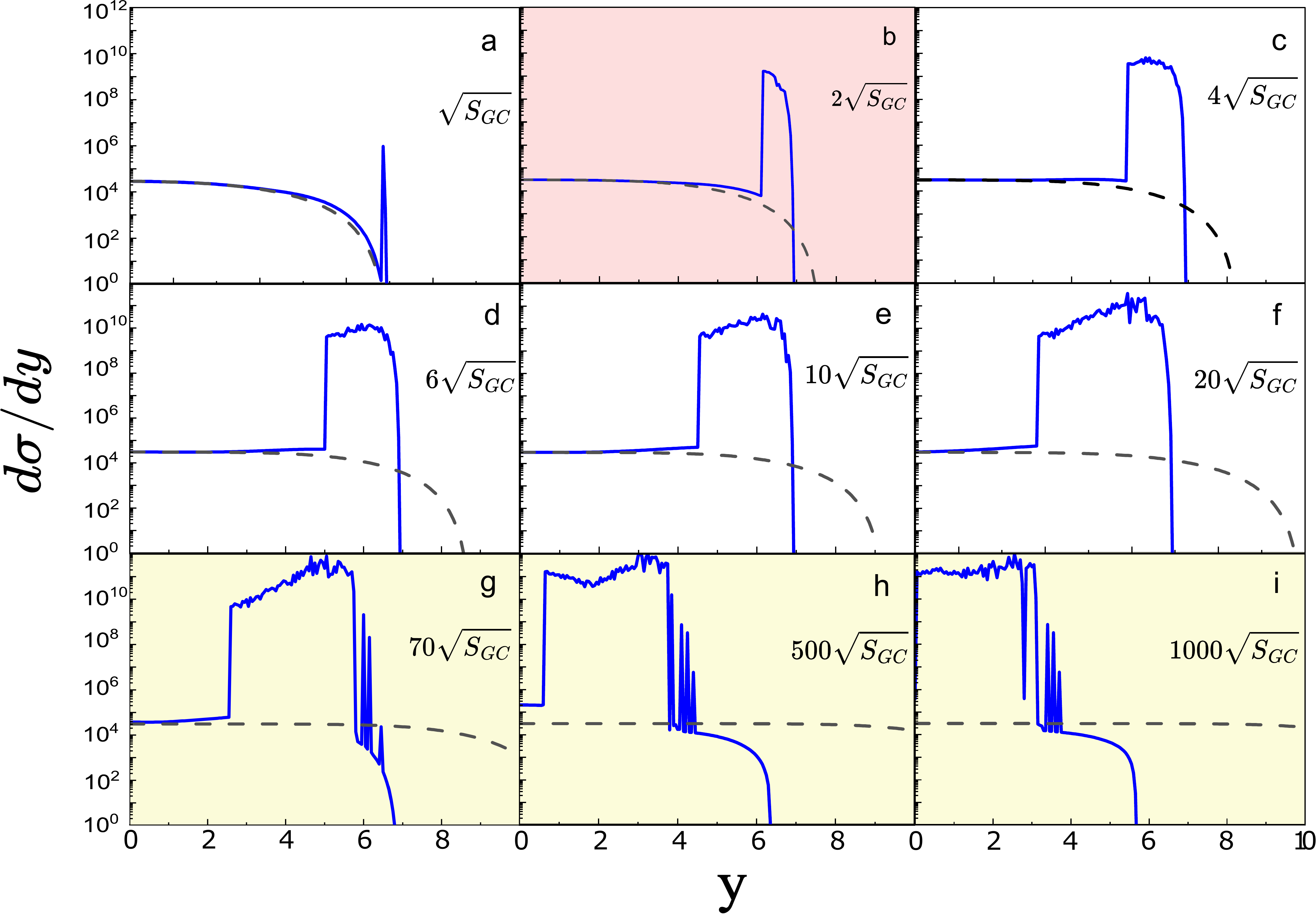}
\centering
    	   	\caption{Inclusive gluon rapidity 
 distribution in the $A-A$ collisions using ZSR equation and input Figure~\ref{fig:1} at different energy $\sqrt{s}$. The panels \textbf{a}–\textbf{i} illustrate different regimes of $\sqrt{s}$, ranging from $\sqrt{s_{GC}}$ up to $1000\sqrt{s_{GC}}$. The~resulting blue curves show the large fluctuations are arisen by GC. The~black broken curves indicate
the results without the QCD evolution.Using astrophysical cases, we estimate that gluon condensation in heavy-nucleus collisions occurs at $\sqrt{s_{GC}} \approx 1.4$ TeV\citep{Zhu2022}. Thus, Figures~\ref{fig:2}b and \ref{fig:2}c correspond to the LHC energy region, where hadronization via a fragmentation model reproduces the results of Begun and Florkowski. Beyond Figure~\ref{fig:2}f, the processes involve ultra-high-energy hadronic collisions producing $\gamma$ rays. They indicate that condensed gluons generate a vast number of mini-jets, whose pion production nearly saturates the available collision energy, thereby forming the BPL structure in the $\gamma$-ray spectra.
}\label{fig:2}
\end{figure}

\begin{figure}[H]
  
\includegraphics[width=0.9\textwidth]{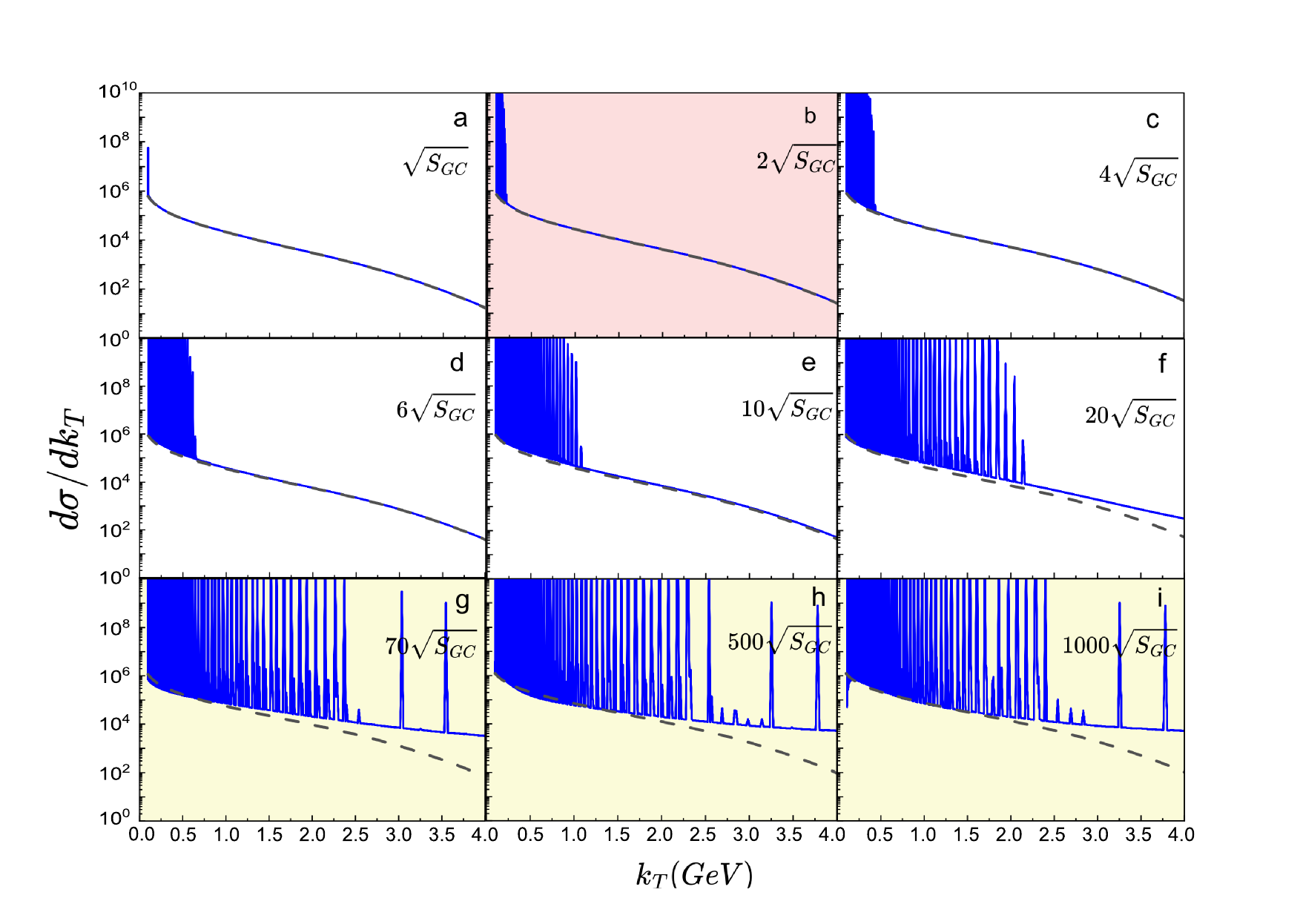}
\centering
    	   	\caption{Similar to
 Figure~\ref{fig:2} but for the $k_T$-distributions of the gluon mini-jets in the $A-A$ collisions. The panels \textbf{a}–\textbf{i} illustrate different regimes of $\sqrt{s}$, ranging from $\sqrt{s_{GC}}$ up to $1000\sqrt{s_{GC}}$.
    }\label{fig:3}
\end{figure}

\begin{figure}[H]
	
		\includegraphics[width=0.8\textwidth]{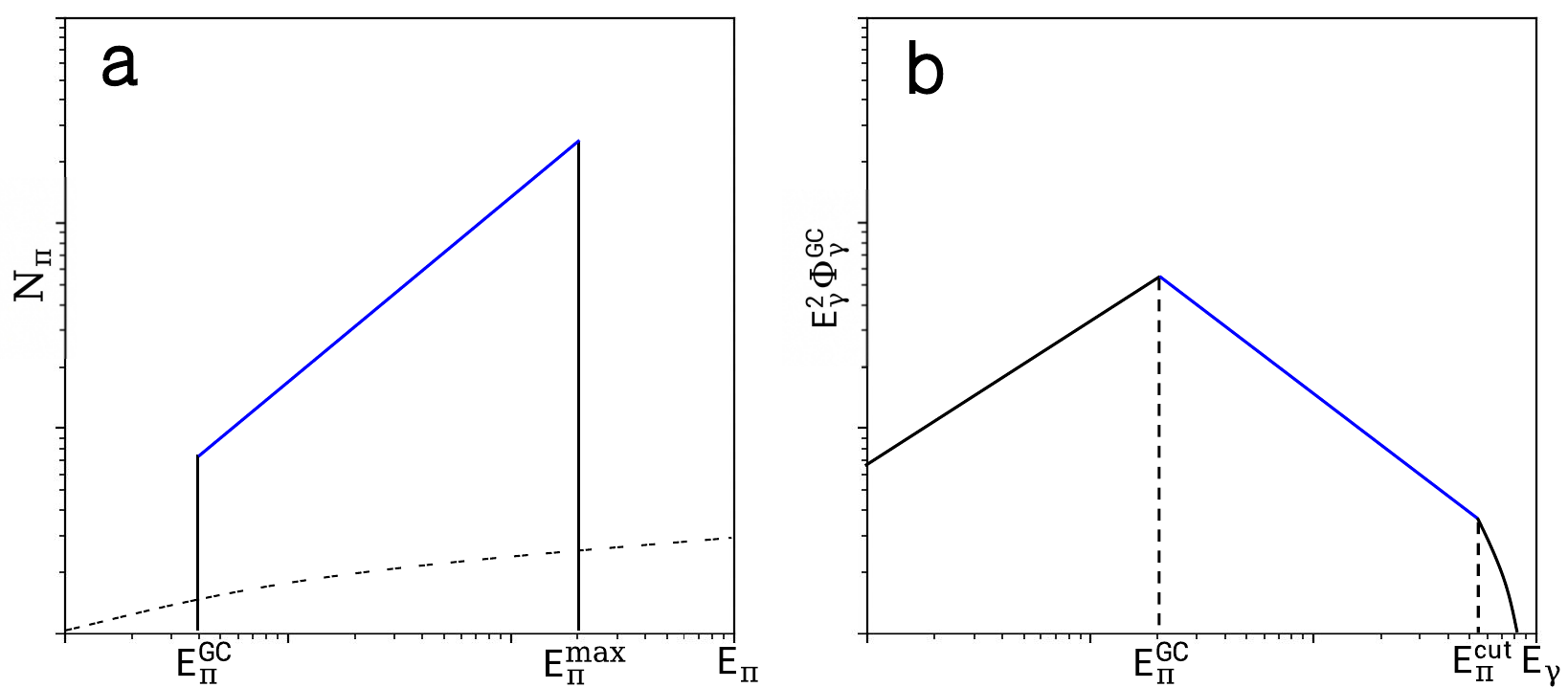}
        \centering
		\caption{(\textbf{a}) The solid curve represents the pion multiplicity $N_{\pi}$ with pion condensation, while the dashed curve represents it without pion condensation. (\textbf{b}) The condensation-spectrum for the VHE gamma-ray spectrum; when the highest proton energy $E_p$ does not reach the condensation threshold $E_p^{max}$, the~gamma spectrum decays exponentially from $E_{\pi}^{cut}<E_{\pi}^{max}$.
        }\label{fig:4}
	
\end{figure}

  Let us first focus on the strong GC region (the yellow areas in Figures~\ref{fig:2} and~\ref{fig:3}). The~$p-A$ collisions are an extremely common phenomenon in
the Universe, and~they are the processes of interest to both astronomical observations and accelerator experiments.
Following Figures~\ref{fig:2} and~\ref{fig:3} which were done using \texttt{origin 10.2.0.188}, we need to consider the hadronization of gluon mini-jets. It is a complex problem. Fortunately, GC provides an ingenious solution to bypass the complexity of the hadronization mechanism \citep{Zhu2020}. We envisage that when a substantial number of condensed gluons at the threshold $x_c$ suddenly participate in the $p-p$ collisions, it inevitably leads to a dramatic increase in the production of secondary pions. Since pions have mass, their yield $N_{\pi}$ is inherently constrained.  In~principle, the~condensed gluons engaging in collisions may simultaneously generate a considerable number of secondary on-mass-shell pions at a given interaction energy; they are capable of saturating all available energy, indicating that nearly all kinetic energies in collisions at the center-of-mass (C.M.) frame are utilized in creating the rest pions. This results in almost no relative momentum for the newly-formed pions, leading to the maximum value of $N_{\pi}$. While the validity of this saturation approximation will be scrutinized by subsequent observed data, adopting this limit allows us to circumvent the complex hadronization mechanism.
Thus, energy conservation and relativistic covariance have
\begin{equation}
E_p+m_p=\tilde{m}_p\gamma_1+\tilde{m}_p\gamma_2+N_{\pi}m_{\pi}\gamma,
    \label{eq:6}
\end{equation}
and
\begin{equation}
    S=(p_1+p_2)^2=2m_p^2+2E_pm_p=(2\tilde{E}^*_p+N_{\pi}m_{\pi})^2,
    \label{eq:7}
\end{equation}
where
$\tilde{m}_p$ marks the leading particle and $\gamma_i$ is the Lorentz factor. Using the empirical relation \citep{Anisovich1985}
$2\tilde{E}^*_p=(1/k-1)N_{\pi}m_{\pi}$ and
$\tilde{m}_p\gamma_1+\tilde{m}_p\gamma_2=(1/k-1)N_{\pi}m_{\pi}\gamma$, where $k\simeq 1/2$ is the inelasticity,
we obtain the relationships between the pion yield $N_{\pi}$, the~proton energy $E_p$, and the pion energy $E_{\pi}$, which exhibits the typical power law (PL), i.e.,~they are the straight lines in double logarithmic coordinates (Figure~\ref{fig:4}a),
\begin{equation}
    \ln N_{\pi}=0.5\ln (E_p/{\rm GeV})+a, ~~\ln N_{\pi}=\ln (E_{\pi}/{\rm GeV})+b,
    \label{eq:8}
\end{equation}
with
$E_{\pi}\in [E_{\pi}^{GC},E_{\pi}^{max}]$, where $a\equiv 0.5\ln (2m_p)/{\rm GeV})-\ln (m_{\pi}/{\rm GeV})+\ln k$
and $b\equiv \ln (2m_p/{\rm GeV})-2\ln (m_{\pi}/{\rm GeV})+\ln k$. 

       Using Equation~(\ref{eq:8}), we obtain
\begin{equation}
    E_p=\frac{2m_p}{m^2_{\pi}}E_{\pi}^2,
    \label{eq:9}
\end{equation}
where $\sqrt{s}\simeq \sqrt{2m_pE_p}$. With~this result, we can determine how much proton energy is needed to produce a specific value of $E_\pi$.

    Equation~(\ref{eq:8}) can lead to striking and distinctive features of GC in proton collisions, provided the GC-threshold energy $E_p^{GC}$
enters the observable range. For~this sake, we examine very high energy cosmic gamma rays.
In many astrophysical processes, protons can be accelerated to unprecedentedly high energies, and~the gamma-ray spectra released by their collisions with other protons or nuclei may carry special GC~information.

\vspace{-6pt}
\vspace{-6pt}

    Substituting the QED formula of $\pi^0\rightarrow 2\gamma$ with Equation~(\ref{eq:8}) into the gamma-ray spectral energy distribution
in the hadronization scenario \citep{Aharonian2012}
\begin{equation}
    \Phi_{\gamma}(E_{\gamma})=C_{\gamma}\left(\frac{E_{\gamma}}{\mathrm{GeV}}\right)^{-\beta_{\gamma}}
\int_{E_{\pi}^{min}}^{\infty}dE_{\pi}
\left(\frac{E_p}{\mathrm{GeV}}\right)^{-\beta_p}N_{\pi}(E_p,E_{\pi})
\frac{d\omega_{\pi-\gamma}(E_{\pi},E_{\gamma})}{dE_{\gamma}},
\label{eq:10}
\end{equation}
where the spectral index $\beta_{\gamma}$ is the photon loss due to absorption in the medium near the source. {The intensity of the proton flux $N_p$ is taken by power law $E_p^{-\beta_p}$ for simplicity. }$C_{\gamma}$ contains the motion factor and the flux dimension; the~following analytical solution is obtained after a simple integration. {This is a BPL (Figure~\ref{fig:4}b)}

\begin{equation}
    E_{\gamma}^2\Phi^{GC}_{\gamma}(E_{\gamma})\simeq\left\{
\begin{array}{ll}
\frac{2e^bC_{\gamma}}{2\beta_p-1}(E_{\pi}^{GC})^3\left(\frac{E_{\gamma}}{E_{\pi}^{GC}}\right)^{-\beta_{\gamma}+2} \\ {\rm ~~~~~~~~~~~~~~~~~~~~~~~~if~}E_{\gamma}\leq E_{\pi}^{GC},\\\\
\frac{2e^bC_{\gamma}}{2\beta_p-1}(E_{\pi}^{GC})^3\left(\frac{E_{\gamma}}{E_{\pi}^{GC}}\right)^{-\beta_{\gamma}-2\beta_p+3}
\\ {\rm~~~~~~~~~~~~~~~~~~~~~~~~ if~} E_{\pi}^{GC}<E_{\gamma}<E_{\pi}^{cut},\\\\
\frac{2e^bC_{\gamma}}{2\beta_p-1}(E_{\pi}^{GC})^3\left(\frac{E_{\gamma}}{E_{\pi}^{GC}}\right)^{-\beta_{\gamma}-2\beta_p+3}
\exp\left(-\frac{E_{\gamma}}{E_{\pi}^{cut}}+1\right).
\\ {\rm~~~~~~~~~~~~~~~~~~~~~~~~ if~} E_{\gamma}\geq E_{\pi}^{cut},
\end{array} \right.
\label{eq:11}
\end{equation}
{We refer} 
 to this as the GC spectrum, which has been used to explain almost a hundred cases of cosmic gamma-ray spectra, including those from supernova remnants (SNRs), pulsars, active galactic nuclei (AGNs), the~Galactic Center, and~gamma-ray bursts (GRBs). Of~course, GC is not the sole explanation for power-law (PL) features in cosmic gamma-ray spectra. For~instance, inverse Compton (IC) scattering also exhibits asymptotic PL behaviour, often described using empirical parametric formulas such as the exponentially cutoff PL:
$dN/dE=N_0 (E/E_0)^{-\Gamma} \exp(-E/E_{cut})$
or the log-parabola: $dN/dE=N_0 (E/E_0)^{-[\alpha + \beta \log(E/E_0)]}$. Therefore, careful comparison reveals their difference from
the GC spectrum can test Equation~(\ref{eq:11}).

\begin{figure}[H]
\centering
\includegraphics[width=0.9\textwidth]{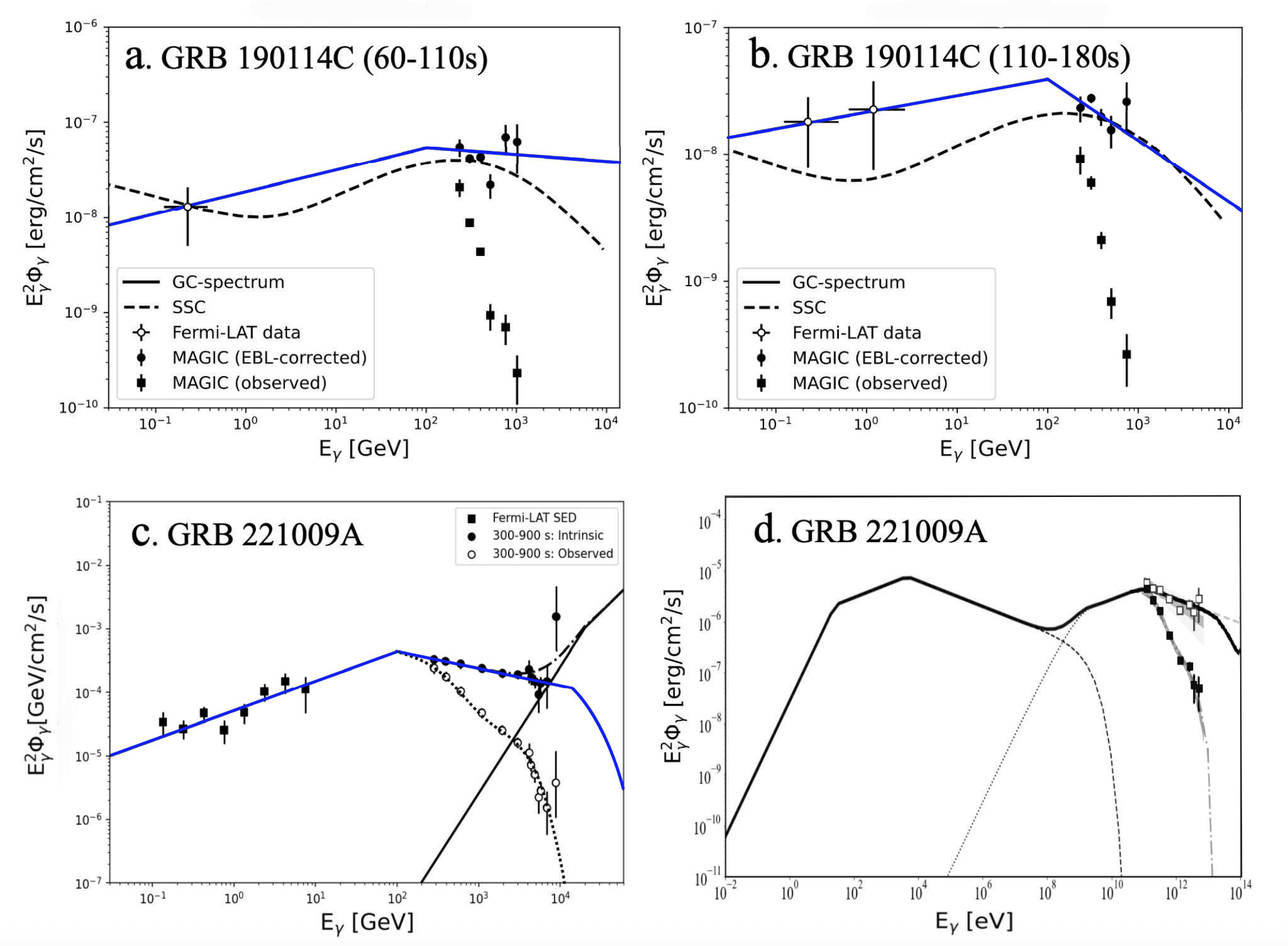}
    \caption{Several GRB gamma spectra. (a, b) Comparisons of the GC spectrum (blue solid curves) fitting GRB 190114C with two leptonic scenarios (IC, dashed curves). (c) The GC spectrum fitting GRB 221009A. (d) An IC model fitting GRB 221009A. The fitting procedure was done using \texttt{iminuit} (\url{https://scikit-hep.org/iminuit/index.html}).}\label{fig:5}
\end{figure}

    We take the GRB spectra at TeV scale as an example. They are of particular interest because of the rarity of such events and the extreme environments. Therefore, careful comparison reveals that their difference from the GC spectrum can test Equation~(\ref{eq:11}). This allows protons to be accelerated into the very high energy region to reveal a more complete GC spectrum. Figure~\ref{fig:5} is a collection of examples where the intrinsic spectra are corrected from earth observations to account for extragalactic background light (EBL) absorption of photons travelling through cosmic space. Figure~\ref{fig:5}a,b show the fits of GRB190114C with two different lepton schemes \citep{Mirzoyan2019}.  The~solid line is the GC spectrum. It seems difficult to judge which model is better. Note that the IC model requires low-energy (KeV) synchrotron radiation as the source of the initial state, its shape being closely related to the TeV spectrum. The~high-energy spectrum predicted by the IC model is governed by the shape of the low-energy spectrum of the synchrotron radiation and~does not necessarily exhibit PL asymptotic behaviour. The~GC spectrum does not have this constraint.
However, a~more complete high-energy spectrum of GRB was recently presented for GRB 221009A, as shown in Figure~\ref{fig:5}c, which clearly favours the the GC spectrum \citep{Zhu2025}. We also note that Foffano et al. present a BPL similar to that of GC using the lepton scheme (Figure~\ref{fig:5}d) \citep{Foffano2024}; however the low- and medium-energy spectra are obviously drawn according to TeV PL, which has no observational fits, making the results~questionable.

    Remember that the broken point of the GC-spectra in this example, $E_{\pi}^{GC}=100$~GeV, is the signature scale at which
GC begins to enter the region of action in heavy nucleus collisions ($p-A$ or $A-A$) \citep{Zhu2022}. We will use it~below.

\section{The BEC of~Pions}\label{sec4}

      A primary goal in ultra-relativistic heavy-ion collisions is to investigate new states of matter under extreme
conditions, especially during the hadronization phase, where pion condensation remains an intriguing and unresolved puzzle. The~collision process can be divided into several stages: initially, the~quark--gluon collision phase, where high-energy quarks and gluons interact, producing numerous mini-jets; these mini-jets then gradually form a quark--gluon plasma (QGP) through radiation and rescattering; finally, as~the system cools to near-critical temperatures, the~QGP undergoes hadronization, resulting in a multitude of hadrons. If~the pion chemical potential approaches the pion mass $(\mu_\pi \rightarrow m_\pi)$ during QGP hadronization, Bose--Einstein statistics predict the macroscopic occupation of pion states with zero momentum $(p\approx 0)$, known as the BEC of~pions.

     Theoretically, the~simplest way to identify such pions is by directly observing their transverse momentum
distribution. However, this method encounters significant experimental and analytical challenges: (i) Particles with very low momentum are difficult to detect accurately. As~momentum nears zero, detector efficiency and precision decrease sharply, making it hard to distinguish condensed pions from ordinary low-energy pions; (ii) The ultra-low momentum region is affected by background contributions from secondary particle scattering, decay processes, and~instrumental noise, which can obscure potential condensation signatures. Therefore, statistical models are necessary to analyse the observed~data.

      We discuss the heavy-ion collisions at the Large Hadron Collider (LHC). The~GC-threshold energy $E_{\pi}^{GC}\approx100$ GeV in heavy nucleus collisions \citep{Zhu2022}.  Equation~(\ref{eq:9}) requires the incident proton to be accelerated to $E_p^{GC}\approx10^6$ GeV or $\sqrt{s_{GC}}\approx1.4~$TeV to allow condensed gluons to enter the collision range. Thus, the~scenario discussed by Begun and Florkowski,
  $\sqrt{s}=2.76$~TeV $Pb-Pb$ collisions, occurs in Figures~\ref{fig:2}b and~\ref{fig:3}b, {which is} at $2\sqrt{s_{GC}}$. Considering that the distribution of pions fragmented from gluon mini-jets will become softer, we can expect that
partially localised low-momentum pions to be observed at a large rapidity in the LHC ($Pb-Pb$) collision energy region. This is exactly what is seen by Begun and Florkowski, but~it is a GC effect and is not related to~BEC.

     Let us explain this new~discovery.

(i) Why do new pions with the GC effect appear at the edge of rapidity at the LHC?  This is a natural consequence of GC kinematics and~is well understood and shown in Figure~\ref{fig:2}b. The~mini-jet that appears at the edge of rapidity consists of a large-$x$ gluon and a small-$x$ gluon, so as the collision energy increases, the~GC peak at small-$x$ will appear first at the edge of rapidity. In collisions of any centrality, large rapidity pions are produced, except that the relative importance of their physical effects varies with centrality. In A-A (small centrality) collisions, pion yields in the central rapidity region dominate because they come from massive fireballs. This is the ideal place to study the QGP. In peripheral collisions (large centrality), although the total number of particles is small, the relative contribution of pions from the large rapidity region becomes more important. It is especially a prime region for studying initial state effects such as CGC and GC, since there is no interference from QGP. Therefore, the data analysed by Begun and Florkowski
falls within the range of Figure~\ref{fig:2}.

(ii) The $k_T$ distribution in Figure~\ref{fig:3}b shows that the jets indeed distribute the low momentum region.
The reason is as follows. The~maxima of the gluon transverse momentum distribution in the GC spectrum, Figure~\ref{fig:1}, are in a similar region $k_c^2\approx Q_s^2$. Therefore, the~main contribution to Equation~(\ref{eq:1}) comes from $(k_T+q_T)^2\approx(k_T-q_T)^2$. Thus, we have $k_T\rightarrow 0$, as shown in Figure~\ref{fig:2}b. Considering that the momentum of each pion after fragmentation softens, these pions have low~momentum.

(iii) The QGP may alter quark--gluon distributions; tracking the evolution of GC in the initial nucleon state requires
minimizing QGP interference during hadronization.
Fortunately, in~the $Pb-Pb$ collisions at $\sqrt{s_{NN}}=2.7$~TeV, the~GC effect appears in the rapidity edge region.
Particles produced in the central rapidity region of heavy ion collisions have weak correlations with particles in the peripheral region. This is due to the following: (a) Mechanism differences. The~central region is dominated by QGP collective effects, while the peripheral region is dominated by fragmentation processes, which have different physical origins; \mbox{(b)
Space-time} separation. Large rapidity differences lead to the interruption of causal connections, resulting in limited correlation lengths.
In fact, the~ALICE collaboration measured particle correlations in the central region $(\vert y \vert < 0.5)$ and forward region $(2.5 < y < 4.0)$ of $Pb-Pb$ collisions at $\sqrt{s_{NN}}= 2.76$~ TeV \citep{Khachatryan2014}. The~results showed that for large $\vert \Delta y\vert$, the~correlation function has no modulation in the $\Delta \phi$ direction, with~correlation coefficients less than 0.05, indicating negligible correlations. The~CMS experiment also reported similar findings \citep{Adam2016}. Therefore, we have omitted the possible QGP~influence.

(iv) Why is the distribution of low-momentum pions jumpy? The reason lies in the GC peak. When a large number of condensed gluons enter the collision region, the~number of pions increases dramatically. This is not only because more gluons in the proton raise the probability of gluon collisions, but~also because gluons are the main component of~pion.

(v) Why do Begun and Florkowski find weak isospin symmetry breaking in low momentum pions?
A large number of pions with a certain energy accumulate in a narrow space during each collision.
Due to the overlap of their wave functions, they may transform into each other during the formation time, i.e.,~$\pi^+ \pi^- \rightleftharpoons 2\pi^0$. However, since $m_{\pi^+} + m_{\pi^-} > 2m_{\pi^0}$ and the lifetime of $\pi^0$ ($10^{-16}$ s) is much shorter than the typical weak decay lifetime of $\pi^{\pm}$ ($10^{-6}$ s - $10^{-8}$ s), the~equilibrium will be broken, and~$\pi^0$ will dominate the secondary processes, allowing us to neglect the contribution of $\pi^{\pm}$. We have observed this effect in the astronomical events.
The above is the case of saturation of pion production in very high energy $p-A$ collisions. The~GC phenomenon in the LHC 2.76 TeV region should occur locally in the large rapidity region, where the above isospin asymmetry is much weaker,
as seen by Begun and~Florkowski.

{(vi) BEC is a macroscopic quantum phenomenon. After~undergoing rapid expansion and cooling, a~heavy-ion collision system may locally enter a “freeze-out” stage characterized by low effective temperature and high particle density.
In this regime, the~pion chemical potential approaches its mass, leading to a divergence in the particle number distribution in momentum space, with~an accumulation particularly near low momentum, especially the zero-momentum state.
GC refers to the possible spontaneous clustering of a large number of gluons inside a hadron into a critical momentum state at extremely high energies (very small Bjorken-$x$). This phenomenon originates from the chaotic dynamics induced by singular structures in nonlinear small-$x$ evolution equations, where strong shadowing/anti-shadowing feedback drives gluons to condensate into the critical momentum mode in the final stage of evolution. }

{There are several possible sources of low-momentum pions, each contributing differently to the correlation function. For~example: 
\begin{enumerate}
    \item Resonance decay. Long-lived resonances produce a large number of low-$p_T$ pions (“halo”) in the final state, which mainly broadens the peak widths \citep{wiedemann1996,csorgo1994}.
    \item Coherent hydrodynamic flow. A~combination of thermal emission and collective flow leads to an enhanced abundance of low-$p_T$ particles. The~spectrum at $p_T \lesssim 0.3~\text{GeV}/c$ appears nearly exponential \citep{heinz2013,gale2013}.
    \item  HBT enhancement due to bosonic symmetry. This results in a single-scale broad peak without $\delta$-spikes \citep{lisa2005,wiedemann1999}.
    \item  BEC emission. Inverse particle cascades or coherent fields generate a “coalescence peak” as $p_T \to 0$ \citep{Begun2015,begun2008}.
    \item GC. As~shown in Figures~\ref{fig:2} and~\ref{fig:3}, the~GC contribution to the correlation function can be approximately represented as a columnar distribution in the large-rapidity region at LHC energies. We plan to investigate this in our future work.
\end{enumerate}}

\section{Discussion}

    Based on the above results, we give the following predictions. The~heavy-ion collision energy at the LHC has been
upgraded from $\sqrt{s_{NN}}=2.76$~TeV to $5.36$~TeV, corresponding to the transition shown in Figures~\ref{fig:2}c and~\ref{fig:3}c,
 where low-momentum pions exceeding $19\%$ may be detected in the forward rapidity side region.
The $p-Pb$ collisions at $\sqrt{s_{NN}}=8.16~$TeV  seem to be a stronger effect; however, it is only half as strong as a $Pb-Pb$ collision of the same energy. The~GC thresholds are $E_{\pi}^{GC}\approx 1~$TeV and $20$~TeV  for the O--O collisions at $\sqrt{s_{NN}}\approx 10~$TeV and $p-p$ collisions at $\sqrt{s_{NN}}=300~$TeV \citep{Zhu2022}, respectively. Thus, we do not discuss them in this work.
Future upgrades, such as the HE-LHC for $p-Pb$ collisions, where the C.M. energy may exceed $20~$TeV, and~FCC-hh or SppC up to $70~$TeV, may allow us to observe how the GC transitions from weak to strong and to unravel its mysteries.
As shown in Figures~\ref{fig:2} and~\ref{fig:3}, with~an energy of $\sqrt{s_{NN}}\approx100$~TeV  or more, the~GC effect will become very strong and dominate the whole rapidity region. We warn that strong gamma rays similar to artificial miniature gamma-ray bursts can be accidentally generated, which may damage the~detector.

    Although GC and BEC in heavy-ion collisions have yet to be confirmed by direct experiments, comparing their
differences is insightful. BEC conditions necessitate that the chemical potential approaches (but does not exceed) the pion mass, $\mu_{\pi} \rightarrow m_{\pi}$. The~BEC phase transition is characterized by a significant number of near-zero-momentum $(p \approx 0)$ pions produced in hadron collisions, with~a phase space density surpassing the BEC critical threshold, $N_{\pi} \lambda_{\pi}^3 > 2.612$. Here, $N_{\pi}$ represents the pion number density and~$\lambda_{\pi} = h / \sqrt{2\pi m_{\pi} k_B T}$ is the thermal de Broglie wavelength. Additionally, condensed pions should display unique correlations. Although~these BEC conditions might arise during QGP hadronization, their experimental verification is highly~challenging.

    In contrast, the GC model predicts low-momentum pions that are independent of the BEC mechanism, requiring comparatively relaxed conditions. However, it requires that the proton's energy exceeds the GC threshold. In~ultra-high energy cosmic ray events, the~unique BPL spectrum of GC can be compared with other models to determine if low-momentum pions dominate. Whether or not the GC effect also satisfies the BEC condition remains a separate question. One possibility is that GC produces enough zero-momentum $(p^2 \approx 0)$ pions to contribute to~BEC.

    Conclusions. In~this work, we present, for~the first time, a~unified explanation for anomalies observed in the cosmic gamma-ray spectrum at astrophysical scales and low-momentum pion clustering at the particle scale in heavy-ion collisions at the LHC, attributing both phenomena to GC. Specifically, (1) by employing energy conservation and relativistic covariance, we circumvent the complexities of the hadronization process and derive an analytical solution for gamma-ray spectra that accurately reproduces the BPL features observed in nearly one hundred astrophysical sources, including supernova remnants and active galactic nuclei; and (2) we demonstrate that the low-momentum pion clustering observed by the ALICE collaboration, previously interpreted as BEC, can instead be understood as a manifestation of GC effects. Importantly, the~formation of GC does not require the stringent conditions necessary for BEC, and~it can be tested in future high-energy collider experiments such as the HE-LHC and FCC-hh. We also caution about the potential generation of artificial gamma-ray bursts in ultra-high-energy collisions, emphasizing the importance of detector protection. This series of findings indicates that GC constitutes a novel and under-recognized structural phenomenon within the Standard Model framework, opening new avenues in particle physics, astrophysics, and~nonlinear dynamics. Not only does this deepen our understanding of the proton's internal structure, but~it also provides a powerful tool for exploring matter behaviour under extreme conditions in the~Universe.

\vspace{20pt}

\noindent \textbf{Appendix}  Small-x Gluon Condensation and the Origin of the Slope Explosion Phenomenon. 

For the benefit of readers unfamiliar with QCD evolution dynamics, we offer the following comment from \texttt{ChatGPT 5.1}.

In the high-energy (small-x) region, the parton distribution function of gluons in proton/nucleon is predicted to rise sharply. This steep increase can cause the slope (derivative) of the distribution to diverge at small x, a phenomenon vividly termed slope explosion. This behavior was first explicitly proposed in the groundbreaking 1983 study by Gribov-Levin-Ryskin (GLR) \citep{Gribov1983}. The GLR team consisted of the renowned Soviet theoretical physicist V. Gribov and his colleagues E. Levin and M. Ryskin (all affiliated with the Leningrad Nuclear Physics Institute at the time). Gribov was a pioneer in strong interactions and Regge theory, while Levin and Ryskin were his students or collaborators. While studying small-x QCD, the GLR team discovered that, according to linear evolution equations, the gluon density would grow as a power law as x decreased, violating the unitarity bound for scattering cross-sections. They thus proposed the parton saturation theory: when the gluon density becomes sufficiently high, newly produced gluons undergo fusion, counteracting the growth and slowing or even halting the unchecked increase of the gluon distribution. In their paper, GLR derived an evolution equation incorporating nonlinear terms (later known as the GLR equation), providing the first quantitative description of the saturation effect caused by gluon recombination in high-density gluon systems. This effectively predicted that the slope of the small-x gluon distribution would explosively increase without nonlinear effects, while the saturation mechanism would flatten the distribution, preventing slope divergence. The GLR paper laid the foundation for small-x QCD saturation theory and is regarded as the first work to explicitly identify the slope explosion problem and propose its solution.

The 1986 study by Mueller-Qiu (MQ) further developed GLR's ideas. The authors were A. Mueller, a professor at Columbia University and a prominent high-energy QCD theorist, and J.W. Qiu, a Chinese physicist (then a postdoctoral researcher in the U.S. and
now the director of Jefferson Lab’s Theory Center). In their paper, Gluon Recombination and Shadowing at Small x, Mueller and Qiu independently derived a nonlinear DGLAP evolution equation similar to GLR's (often referred to as the GLR-MQ equation) \citep{Mueller1986}, which included a gluon fusion term. They also noted that the linear DGLAP equation would lead to a steep rise in the gluon distribution function $G(x, Q^2)$ at small x, violating the physical limit of power-law growth. However, with the addition of the fusion term, the gluon distribution exhibited saturation or shadowing effects at high densities, significantly slowing the growth slope. The MQ work corroborated GLR's findings: both viewed the uncontrolled slope explosion as a phenomenon requiring suppression through gluon recombination/saturation mechanisms and provided mathematical descriptions. The Mueller-Qiu paper cited GLR's work, solidifying the credibility of small-x parton saturation theory and widely disseminating it in the Western high-energy physics community.

In the 1990s, experiments at the HERA accelerator measured the steep rise of the proton's deep inelastic scattering structure function $E_2$ at small x, confirming the theoretical prediction of slope explosion: $F_2$ increased rapidly with $Q^2$ as $x\rightarrow 10^{-4} -10^{-5}$, with a large logarithmic slope. This spurred further development of saturation theory. In 1994, L. McLerran and R. Venugopalan proposed the famous Color Glass Condensate (CGC) model, depicting small-x gluons in high-energy nuclei forming a high-density saturated state \citep{McLerran1994}. Despite the term condensate, CGC refers to a coherent, classical, and uniform saturated state of gluons, not a strict quantum mechanical condensate. However, this concept vividly reflects that, in the small-x limit, gluons fill phase space with extremely high occupancy, forming a collective state resembling a condensate. The CGC effective theory was systematically developed around 1999 by Iancu, Ferreiro, Jalilian-Marian, Kovner, and others (via the JIMWLK equation and Kovchegov's BK equation). These nonlinear evolution equations also stemmed from the GLR-MQ framework, incorporating multiple scattering and fusion to constrain gluon distribution growth. The Balitsky-Kovchegov (BK) equation is a well-known result in the CGC framework, predicting that the gluon distribution approaches a "black disk" saturation limit at small x, where the distribution function no longer increases indefinitely. However, it is worth noting that traditional CGC/BK evolution yields saturation solutions that are typically flattened distributions, without sharp peaks or divergent derivatives. In other words, CGC describes a saturation equilibrium rather than a true "slope explosion" spike.

The idea of slope explosion as a signal of gluons condensing into a peak at a specific momentum was elevated by the work of Chinese scholar W. Zhu and his team in the 21st century. Zhu (a professor at East China Normal University) and his collaborators, began refining the GLR-MQ equations in 1999, introducing more complete higher-order loop contributions and stochastic coherence effects (they named the developed evolution equations the Zhu-Ruan-Shen (ZRS) equation and Zhu-Shen-Ruan (ZSR) equation) \citep{Zhu2016}. In a 2008 paper, Zhu's team first reported a striking feature of the solution: as the evolution progressed, the gluon distribution exhibited a pronounced peak at a critical transverse momentum, as if a large number of gluons were converging at that momentum. This phenomenon was termed GC by the authors, suggesting that gluons in a high-density, small-x environment undergo behavior analogous to Bose-Einstein condensation. Mathematically, this manifests as a delta-function-like spike in the distribution function, where the derivative of the gluon distribution with respect to momentum diverges at the critical momentum $k_c$. Here, slope explosion acquired a new physical meaning: the derivative of the gluon distribution with respect to momentum approaches divergence at the peak, with the curve rising steeply as if exploding before sharply descending. In 2016-2017, Zhu and his team confirmed the existence of this gluon condensation through more precise ZSR evolution equations. They found that the antishadowing oscillations introduced by nonlinear terms caused violent fluctuations in the gluon distribution, leading to the formation of sharp peaks at critical points. The authors quantitatively described these peaks as $\delta$-function-like distributions and defined the corresponding critical condensation momentum $k_c$. This marked a shift in the understanding of slope explosion: it was no longer merely a pathological growth to be suppressed but a potential signal of a new state of matter¡ªif experiments observed an anomalous peak-like accumulation of gluons at a specific momentum, it would directly confirm gluon condensation.

In summary, GLR (1983) first identified the steep rise (slope explosion) of small-x gluon density as a potential issue and proposed saturation as a solution. Mueller-Qiu (1986) followed, theoretically corroborating and refining this idea. Both primarily used terms like saturation, recombination, or shadowing, emphasizing the suppression of gluon growth toward flattening. The CGC theory emerged around 2000; although its name includes condensate, it essentially describes a saturated equilibrium state where high-density gluons occupy phase space like a Bose gas. It was not until the work of Zhu et al. (2008, 2016) that the term gluon condensation was endowed with a stricter meaning: they predicted the appearance of a sharp peak in the gluon distribution with an infinite slope -- a true slope explosion. In terminology, GLR/MQ tended to describe saturation, while Zhu's team directly adopted condensation to characterize the peaked distribution; CGC lies between the two, describing a saturated state without sharp peaks. Chronologically, GLR was the pioneer, followed by MQ; CGC theory inherited and developed GLR's ideas; Zhu et al.'s condensation concept was a further prediction, later in time but more radical in concept. They explicitly cited and extended the GLR-MQ equations (calling them the GLR-MQ-ZRS equations) and noted that the conventional BK saturation equation does not produce condensation solutions, highlighting the need for their improvements. Thus, slope explosion evolved from a theoretical warning to a conjecture about a new condensed phase, reflecting a conceptual progression from saturation to condensation.

\end{document}